# Integration of Physical Equipment and Simulators for On-Campus and Online Delivery of Practical Networking Labs

KA CHING CHAN

Department of Computer Science and Information Technology,
La Trobe University, Bendigo, Victoria 3552, Australia.
E-mail: `ka.chan@latrobe.edu.au`

This paper presents the design and development of a networking laboratory that integrates physical networking equipment with the open source GNS3 network simulators for delivery of practical networking classes simultaneously to both on-campus and online students. This transformation work has resulted in significant increase in laboratory capacity, reducing repeating classes. The integrated platform offers students the real world experience of using real equipment, and the convenience of easy setup and reconfiguration by using simulators. A practical exercise in setting up an OSPF/BGP network is presented as an example to illustrate the experimental design before and after the integration of GNS3 simulators. In summary, we report our experiences with the integrated platform, from infrastructure, network design to experiment design; and the learning and teaching experiences of using GNS3 in classes.

## 1. Introduction

The undergraduate Information Technology program at La Trobe University is a three year program. In the networking stream, there are one level 2 (Computer Networks) and two level 3 subjects (Internetworking, and IT and Information Security). The IT course is highly industry relevant [1], and hands-on labs are integral parts in all these networking subjects. In the past, networking experiments were delivered in the networking laboratory using physical networking equipment. Due to the high cost of equipment and limited space, repeating lab sessions were necessary. The networking laboratory supports not only teaching, but also all the project and research activities, such as cloud robotics [2], broadband access networks [3], communications [4, 5], and application development [6, 7].

As the direction of the University's learning and teaching strategy has been gradually steered towards a digital future, the flipped classroom model and blended learning have been adopted in increasing number of subjects. We have previously attempted to develop a virtual networking lab using the then open source software Vyatta [8]. In 2012, Vyatta was acquired by Brocade and renamed Brocade virtual router [9] and is no longer available for free. In 2013, an independent group started a fork and continued the community version under the name VyOS [10].

This paper presents our current effort in meeting the University's strategic plan and modernising the delivery of networking hands-on labs. To achieve this goal, we have modified the layout of two laboratories, redesigned the network, added a virtualisation server (VMWare ESXi) [11] and two network attached storage (NAS) servers (QNAP TS-420U) [12], installed multiple network simulators (GNS3) [13] as virtual machines, integrated GNS3 with the existing Cisco physical networking equipment, and redesigned the laboratory exercises. Due to the restructure, laboratory capacity has been significantly increased. Students can now perform their experiments in any of the networking lab and the PC labs in the department, or remotely from other campuses or their homes. The tasks involved are described in the following sections. This paper describes the transformation work involved in the process, and reports our experiences in delivering practical classes using the new approach.

## 2. LABORATORY NETWORK DESIGN

In this section, we will describe the laboratory layout, network setup, and the networking equipment and software that support the delivery of hands-on practical exercises.

### 2.1. Networking Laboratory

There are three laboratories in the department, and its floor plan is shown in Fig. 1. The networking laboratory, on the left, has a capacity for up to four groups of three students to carry out their hands-on labs concurrently using the physical equipment. At the back is the server room that hosts a full rack of networking equipment to support all the teaching and research activities in networking. Due to the limited hardware and space available, two to three repeating sessions per week were required for each subject. The other two computer labs (the PC lab and Mac lab), are within proximity to the networking lab, as shown in the middle. The layouts of these two labs have been modified to include small movable round tables for group work. The objective was to use GNS3 to extend the capacity of the networking lab, allowing the whole class of students to work on the same lab session simultaneously in either the networking lab, PC lab, Mac lab, or remotely online. By using GNS3, a single lab session will increase the capacity to host over 50 on-campus students. Additional remote students, from home or other campuses, can also join in and work collaboratively online with the on-campus students.

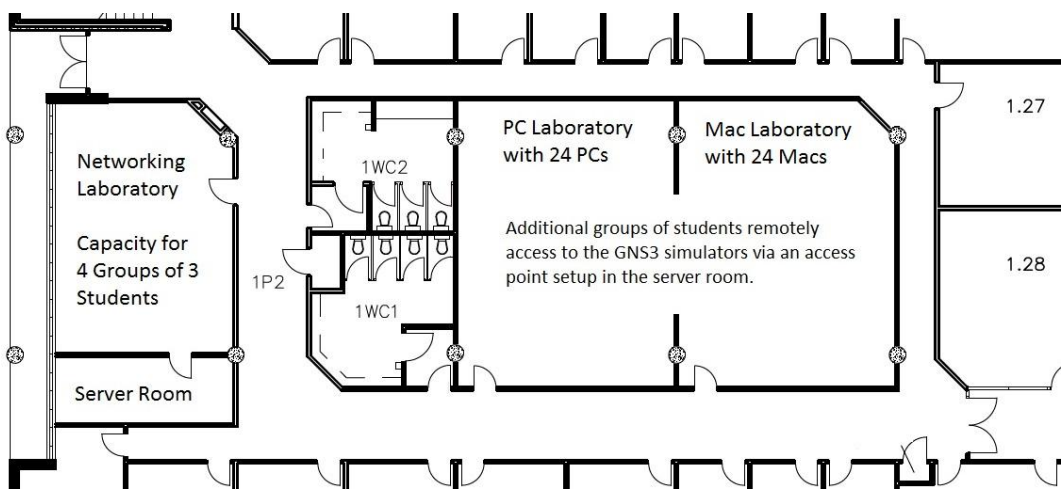

Fig. 1. Department floor plan

### 2.2. Network Setup and Networking Equipment

The network diagram is given in Fig. 2. The Internet traffic is handled via a Cisco 7206VXR router located in a carrier-neutral data center in Sydney, multi-homed to multiple tier-1 carriers. A full class C public IPv4 and a /48 IPv6 have been allocated to our exclusive use. These public IP addresses are mapped to our servers through a GRE tunnel. The advantage is that traffic going in and out of our network is separated from the campus network. Therefore, the security of the campus network will not be compromised due to our teaching and research activities, which may produce security vulnerabilities intentionally from time to time.

The laboratory network is divided into a demilitarised zone (DMZ), as shown in the top left corner of Fig. 2, and a private zone, as shown in the bottom half of Fig. 2. A number of servers that require public IP addresses, including a public web server, a Wikipedia server, a VoIP server, and additional servers supporting research and student projects are located in the demilitarised zone. The networking systems used for student experiments are located in the private network zone. In terms of physical equipment, the networking laboratory is equipped with four racks of routers and switches. Each rack is mounted with the following physical networking equipment:

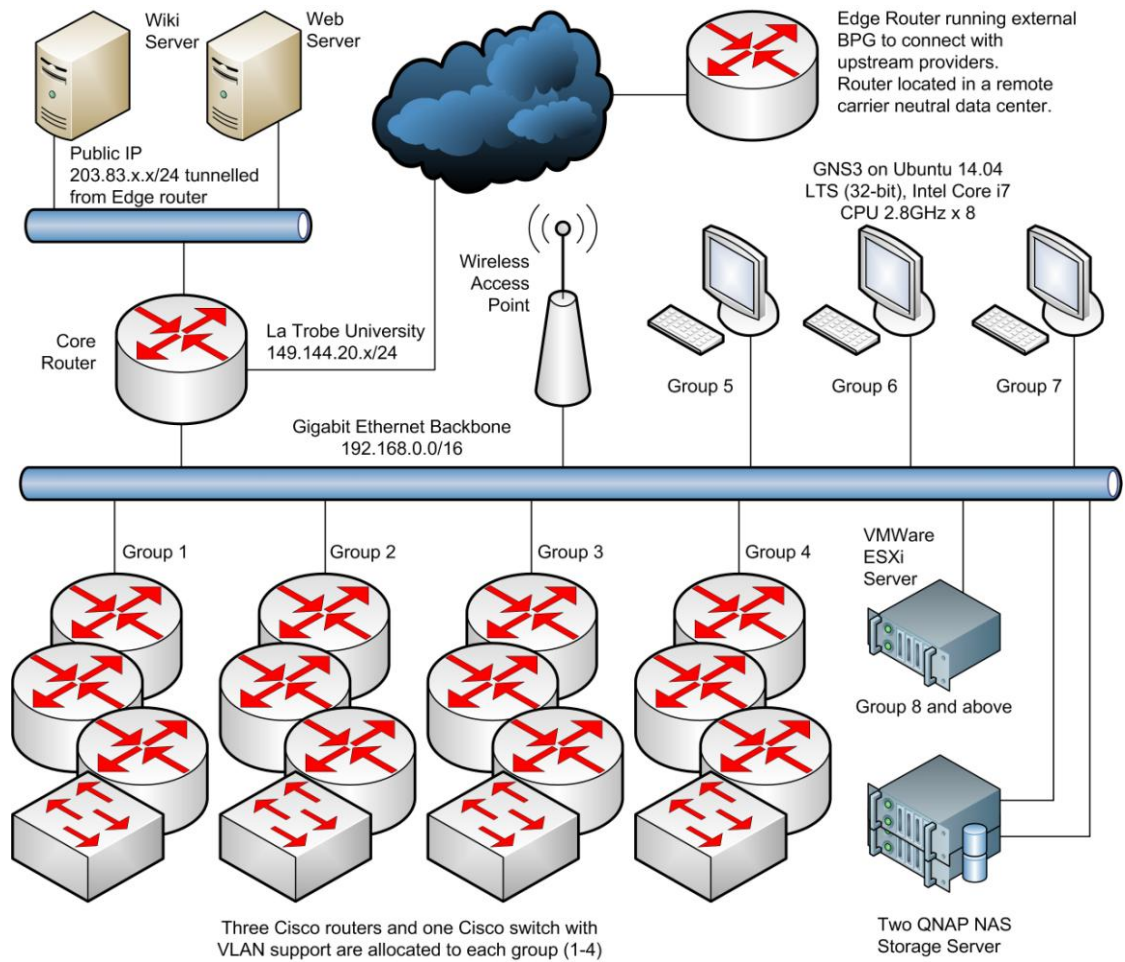

Fig. 2. Network design for the networking laboratory

- A Cisco 2801 router (4 Fast Ethernet ports and 1 ADSL port)
- A Cisco 2610 router (5 Ethernet ports)
- Two Cisco 1841 routers (2 Fast Ethernet ports)
- A Cisco Catalyst 2960 switch (24 Fast Ethernet ports, 2x1G Uplink ports)
- Two Cisco Catalyst 3750G switches (24x1G PoE ports, 4x1G SFP)

These existing physical systems provide students the opportunity to carry out some of their experiments using real systems, a valuable experience especially for beginners in networking. In the following sections, we will discuss how we increased our laboratory capacity by integrating the existing networking equipment with a virtualisation platform running the network simulator GNS3.

## 2.4. Virtualisation Platform and Network Simulator

In the first phase of this work, we investigated the feasibility of using GNS3 for teaching on three old PCs with Intel Core i7 CPUs running Ubuntu OS (Desktop version 14.04 LTS) [14]. We allocated these three GNS3 servers (version 1.2.3) to student group 5 to 7, as shown in Fig. 2. Once we were confident about the performance and reliability of them, we implemented the same GNS3 simulators on a virtualisation platform that includes a VMWare ESXi server and two QNAP storage servers. Setting up multiple virtual machines on a VMWare made the management of simulators simple and easy. Expansion and maintenance tasks have become mostly copying files, taking snapshots, and backing up, all performed on one machine.

Each GNS3 simulator is setup as one VMWare virtual machine. All virtual machines are stored in one of the two QNAP rack-mounted storage servers, connected through iSCSI to the VMWare ESXi server. Each group (Group 8 and above) of students is allocated a VMWare virtual machine with GNS3 installed on Ubuntu. All the GNS3 simulators are connected to the same backbone, which is a gigabit Ethernet IP network 192.168.0.0/16, via the virtual switch of the VMWare ESXi server, as shown in the middle part of Fig. 2. One key design feature of our network is that all the physical systems, GNS3 on physical servers, and GNS3 on VMWare virtual servers are connected to the same backbone. A Wifi access point has been setup to link all three rooms to the backbone. As shown in later sections, the experiments have also been redesigned in such a way that all students can work on the same experiments and achieve the same learning outcomes regardless of the systems they use. This backbone also enables us to integrate other non-Cisco systems such as VyOS [10] and pfSense [15] in the experiments. VyOS is a community fork of the Linux based router Vyatta that provides software-based routing, firewall, and VPN functionality; and pfSense is becoming a popular open source network security solution. As non-Cisco technology, both VyOS and pfSense can be setup as virtual machines offering our students the opportunity to learn all the important networking concepts in a vendor-neutral environment.

The students in Group 8 and above access the GNS3 simulators using Teamviewer [16], a remote access and screen sharing software for online meeting, web conferencing, and online collaboration. A Teamviewer online meeting can hold up to 25 participants (a lot more than required), and allow participants to communicate via chat, video conferencing or voice over IP (VoIP). Teamviewer was chosen because public IP addresses are not required for remote access from outside. For on-campus students, they can access their GNS3 from the PC laboratory or the MAC laboratory, and collaborate with their online group members at the same time using Teamviewer.

## 3. DESIGN OF HANDS-ON LAB EXPERIMENTS

The original lab experiments have been designed with group collaboration in mind. Networking requires working relationship with other networking engineers, e.g. providers, customers, and so on. Hands on experiments, complementing theoretical concepts presented in lectures, play an integral part in all the subjects in the networking stream. The lab experiments cover the following topics: VLAN, Static routes, access lists, IPv6, RIP, OSPF, BGP, route redistributions, route maps, VPN, and so on.

To highlight how GNS3 helps increase the capacity of the laboratory and enable online delivery, both the original and redesign of an example exercise are presented.

### 3.1. Original OSPF/BGP Setup

This section presents one of the practical classes for the subject CSE3INW Internetworking. In this exercise, students are required to setup an internal network running OSPF within an Autonomous System (AS) using two routers, a primary router (PR) and a secondary router (SR). The interior network is then connected to the outside world, via two other ASs using the external routing protocol BGP. The OSPF networks are then redistributed into the BGP network and vice versa. This scenario is a typical multi-homed network running BGP as external routing protocol and OSPF as internal. The details required for the configurations are shown in Fig. 3 and Table 1. The laboratory tasks are listed below:

1. Configure each router interface with the appropriate IP address.
2. Configure BGP and establish sessions with the appropriate peers.
3. Enter the appropriate commands so that BGP advertises all directly connected networks.
4. Ensure that your router is learning routes from your peers.
5. Ensure that other routers are learning routes that are being advertised by your router.
6. Connect your workstation into your switch and configure it with an appropriate IP address. Verify that you can ping another group's workstation (or at least one of their "internal" router interfaces).
7. Try using traceroute and see if you can discover what path your packets are taking between workstations.
8. Try to identify the AS Path that your router uses for each Autonomous System.

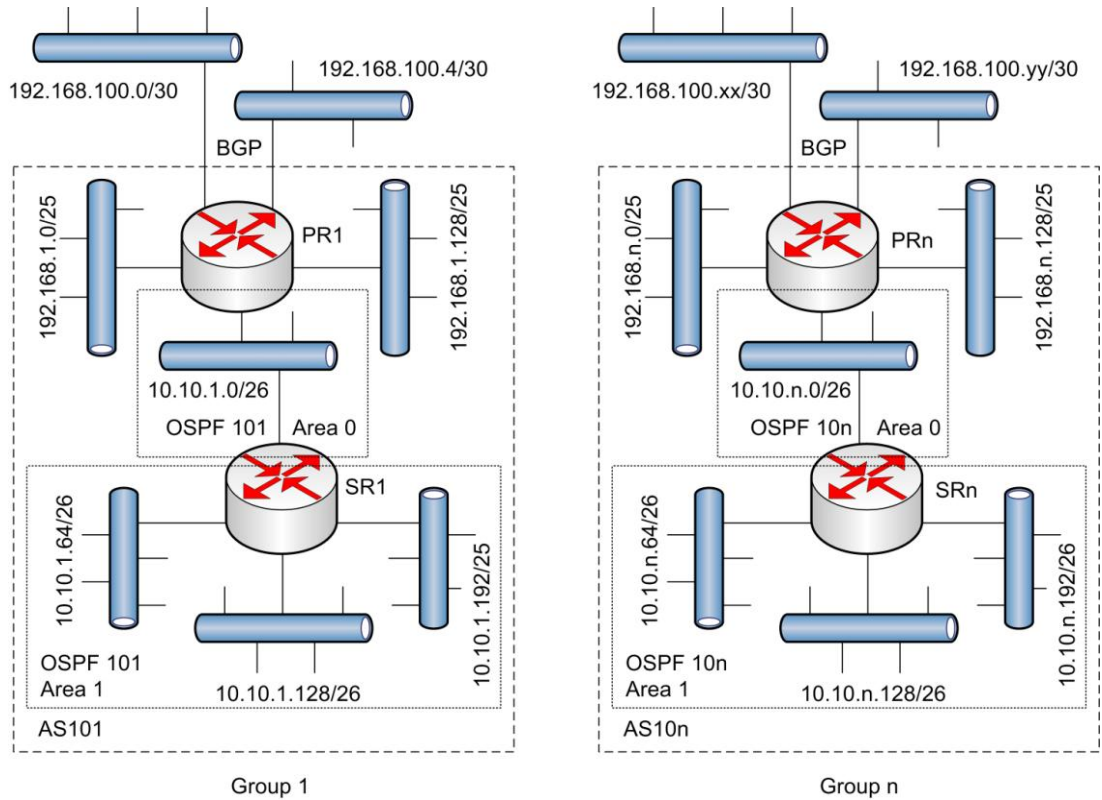

Fig. 3. Network diagram for the OSPF/BGP exercise

Table 1.  Interface IP addresses of the original laboratory setup

| **Network** | **Group 1 AS101** | **Group 2 AS102** | **Group 3 AS103** | **Group 4 AS104** |
|---|---|---|---|---|
| PRn, F0/0, BGP Peer 1 | 192.168.100.1/30 | 192.168.100.2/30 | 192.168.100.9/30 | 192.168.100.10/30 |
| PRn, F0/1, BGP Peer 2 | 192.168.100.14/30 | 192.168.100.5/30 | 192.168.100.6/30 | 192.168.100.13/30 |
| PRn, F0/3/0, VLAN 1 | 192.168.**1**.1/25 | 192.168.**2**.1/25 | 192.168.**3**.1/25 | 192.168.**4**.1/25 |
| PRn, F0/3/0, VLAN 2 | 192.168.**1**.129/25 | 192.168.**2**.129/25 | 192.168.**3**.129/25 | 192.168.**4**.129/25 |
| PRn, F0/3/1, OSPF Area 0 | 10.10.**1**.2/26 | 10.10.**2**.2/26 | 10.10.**3**.2/26 | 10.10.**4**.2/26 |
| SRn, E1/0, OSPF Area 0 | 10.10.**1**.1/26 | 10.10.**2**.1/26 | 10.10.**3**.1/26 | 10.10.**4**.1/26 |
| SRn, E1/1, OSPF Area 1 | 10.10.**1**.65/26 | 10.10.**2**.65/26 | 10.10.**3**.65/26 | 10.10.**4**.65/26 |
| SRn, E1/2, OSPF Area 1 | 10.10.**1**.129/26 | 10.10.**2**.129/26 | 10.10.**3**.129/26 | 10.10.**4**.129/26 |
| SRn, E1/3, OSPF Area 1 | 10.10.**1**.193/26 | 10.10.**2**.193/26 | 10.10.**3**.193/26 | 10.10.**4**.193/26 |

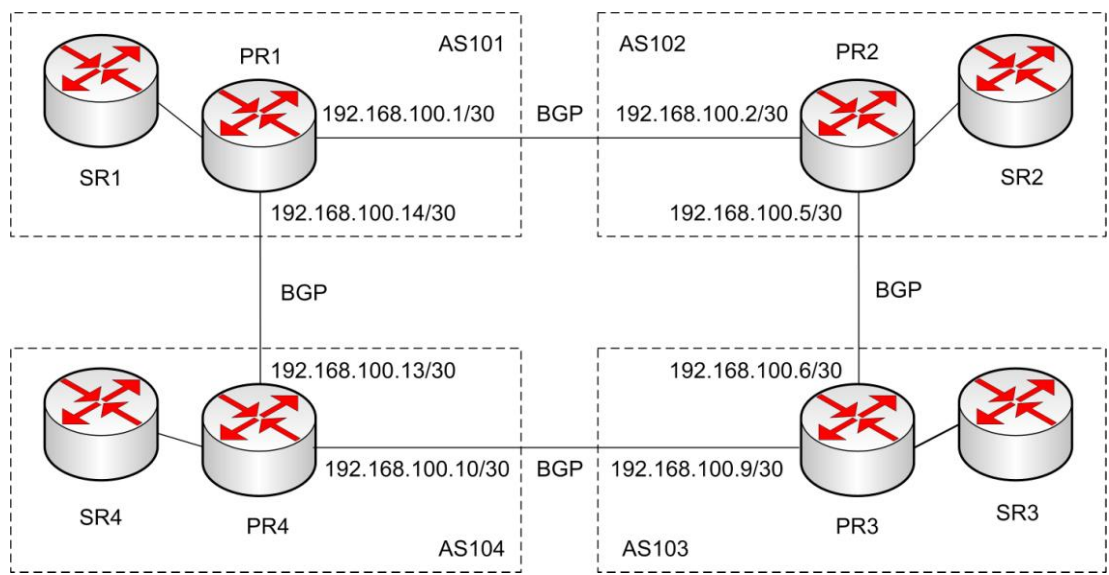

Fig. 4. BGP peer details for the OSPF/BGP exercise

9. Configure your primary router. Ensure that your BGP sessions are correctly established.

10. Configure each interface on your secondary router with the appropriate IP address.

11. Configure OSPF on both your primary and secondary routers. Ensure that passive-interface statements are specified for your BGP peering interfaces.

12. Enter the appropriate commands so that BGP advertises all networks that have been learned via OSPF.

13. Configure OSPF so that it redistributes all routes learned via BGP.

14. Ensure that other routers are learning routes that are being advertised by your router.

15. Connect your workstation into your switch and configure it with an appropriate IP address. Verify that you can ping another group's workstation (or at least one of their "internal" router interfaces).

16. Try using traceroute and see if you can discover what path your packets are taking between workstations.

17. Record your router's running configuration, routing table and BGP table.

Each group has to perform the above tasks independently. Due to the design of the experiment, no group can complete its tasks until all other groups had their networks successfully configured. This can be an issue as students do not have the same standard nor learn at the same pace. This also puts pressure on the lecturers or demonstrators as they are required to assist and troubleshoot, and to ensure all networks work properly so students can perform testing, and collect information and data for their reports. In the next section, we will describe how GNS3 can help overcome these issues and provide a scalable environment to meet resource demands and enable the delivery of practical classes online.

## 3.2. Redesign of the OSPF/BGP Setup

This section presents the redesign of the previous laboratory exercise for multi-campus and online delivery. To eliminate the dependency of each other and enable students to learn at their own paces, three BGP/OSPF demo networks are pre-configured in GNS3 for all student groups to connect to. Students are required to setup their own networks to connect with these demo networks, and investigate how the routing protocols work. Previously each group set up its BGP peers with two other groups, and therefore, they depend on each other. The main modification is that all groups will now connect to the same BGP neighbours on demo networks. These demo networks are pre-configured before the lab sessions and are on the same backbone as the physical equipment and the GNS3 simulators. Fig. 5 shows the three demo networks and the network of group *n*. Table 2 shows the IP addresses and the AS number that group *n* should use. In this new design, each group can just focus on getting its own networks to work, without hindrance from each other.

The first four groups, $n = 1$ to 4, carry out their lab tasks on the physical equipment. The other groups, $n \geq 5$, carry out their work on GNS3 simulators. The use of GNS3 overcomes the capacity issue, and more groups can be accommodated either in the PC lab and Mac lab close by, Fig. 1. Students in these rooms and other off-campus students can connect to the GNS3 servers via Teamviewer. Fig. 6 shows a GNS3 screenshot of the lab exercise.

## 4. LEARNING AND TEACHING EXPERIENCE

The simulator GNS3 was first introduced in a final year undergraduate subject CSE3INW Internetworking. Students taking this subject have already developed basic skills in configuring Cisco routers and switches from the fundamental networking subject CSE2CN Computer Networks. They are familiar with Cisco hardware, basic routing and cabling. Each week two groups have been chosen to run their lab sessions on GNS3. Before they started their work, a ten minute introduction and demonstration on using GNS3 was given by the lecturer. They were shown how to create a new GNS3 project, set up networks, bring up consoles, connect devices, and capture packets using Wireshark [17]. A short demonstration like this was sufficient for all students due to the intuitive design of GNS3. An interesting finding was that once a group tried using GNS3 for their labs, GNS3 would become their preference for future lab sessions.

Students found GNS3 very user-friendly and easy to use. Also, GNS3 displays the network diagram and allows students to access any device console easily by clicking on the device icons. For larger networks in particular, this feature helps students gain a high level understanding of the network much easier and also allow them to switch back and forth from device to device very easily during configuration. In general, students reported that GNS3 enables them to learn and complete their exercises more efficiently; and they can focus on learning new materials rather than troubleshooting side issues like cabling.

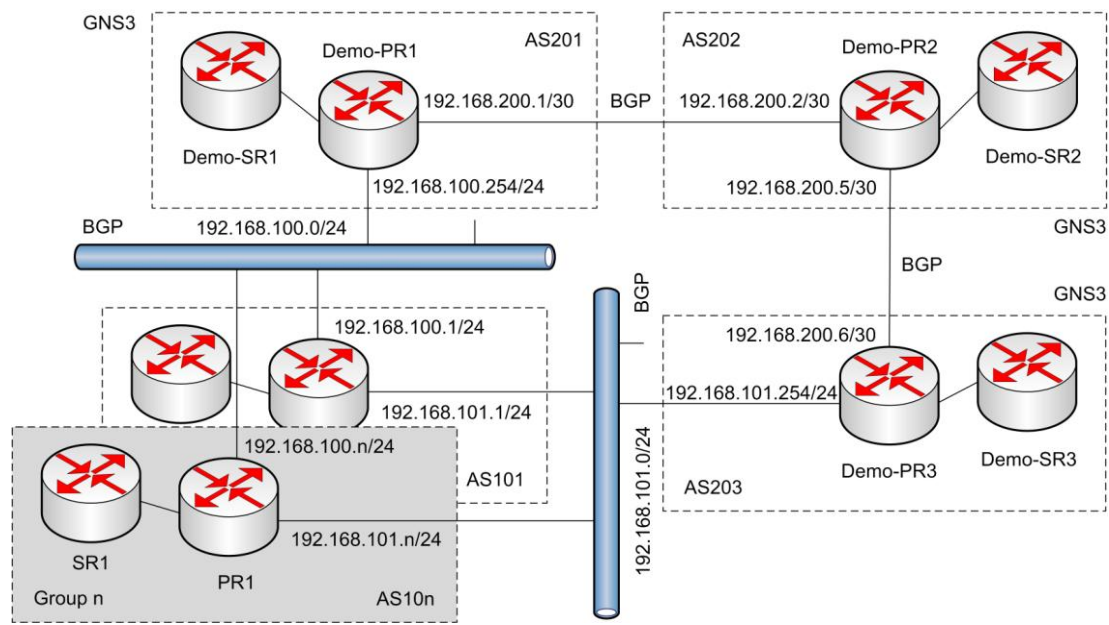

Fig. 5. Redesigned network diagram for the OSPF/BGP exercise

Table 2. Interface IP addresses of the redesigned laboratory setup (**n** = group number)

| Network | Group n AS10n | BGP Neighbour (Common to all groups) |
|---|---|---|
| PRn, F0/0, BGP Peer 1 | 192.168.100.**n**/24 | 192.168.100.254/24, AS201 |
| PRn, F0/1, BGP Peer 2 | 192.168.101.**n** /24 | 192.168.101.254/24, AS203 |
| PRn, F0/3/0, VLAN 1 | 192.168.**n**.1/25 | n/a |
| PRn, F0/3/0, VLAN 2 | 192.168.**n**.129/25 | n/a |
| PRn, F0/3/1, OSPF Area 0 | 10.10.**n**.2/26 | n/a |
| SRn, E1/0, OSPF Area 0 | 10.10.**n**.1/26 | n/a |
| SRn, E1/1, OSPF Area 1 | 10.10.**n**.65/26 | n/a |
| SRn, E1/2, OSPF Area 1 | 10.10.**n**.129/26 | n/a |
| SRn, E1/3, OSPF Area 1 | 10.10.**n**.193/26 | n/a |

As GNS3 is integrated with Wireshark, students can capture packets on any links by simply making a few clicks. Capturing packets on physical equipment is usually more complex as PCs are sometimes required to connect to the same multi-access network. At the end of a lab session, students can easily collect the required information, such as Cisco configuration files, routing tables, and Wireshark captures, and email to themselves or save to their cloud accounts for their write-ups. GNS3 also allows students to save their networks as project files; therefore, it is very easy for students to go back to any step or redo their exercises at later times if needed.

All the GNS3 simulators have been setup with Internet access. Online students can form groups with on-campus students to work collaboratively by using the remote access and web conferencing application Teamviewer. The screen sharing feature of Teamviewer enables remote students to see the same screen as the on-campus students. Students have reported positive feedback on this feature.

Prior to using GNS3, repeating lab sessions had to be run for each networking subject, requiring frequent changeovers from one lab to another multiple times in a week. Switching labs back and forth involve cabling and modifications made to the routers and switches, creating extra work in the preparations. Adding GNS3 simulators increased the lab capacity significantly and repeating lab sessions can be eliminated.

With the extra capacity that GNS3 can easily provide, most of the lab exercises have been redesigned. As shown in the OSPF/BGP example, the lab exercises are now centered around a common backbone that all student groups are required to connect their networks to and communicate to one or more pre-configured demo networks. The preparation for a lab session has become very simple. All is needed from a lecturer is to copy GNS3 project files that contain the whole pre-configured networks to the GNS3 workstations. The redesigned lab setups enable students to work at their own paces and collect information without relying on other groups to have their networks successfully configured.

In summary, GNS3 significantly reduces workloads in lab preparations, and enables student groups to learn at their own paces at any location, and achieve the same learning outcomes. We have also found that GNS3 is intuitive and very easy to learn. Based on our experience, running GNS3 on Ubuntu is very stable and serves the purpose very well. GNS3 works well for fundamental as well as advanced and complicated networking labs. The technology is mature and now is the right time to consider implementing GNS3 in any networking labs. At the time of this writing, a newer version of GNS3 has become available.

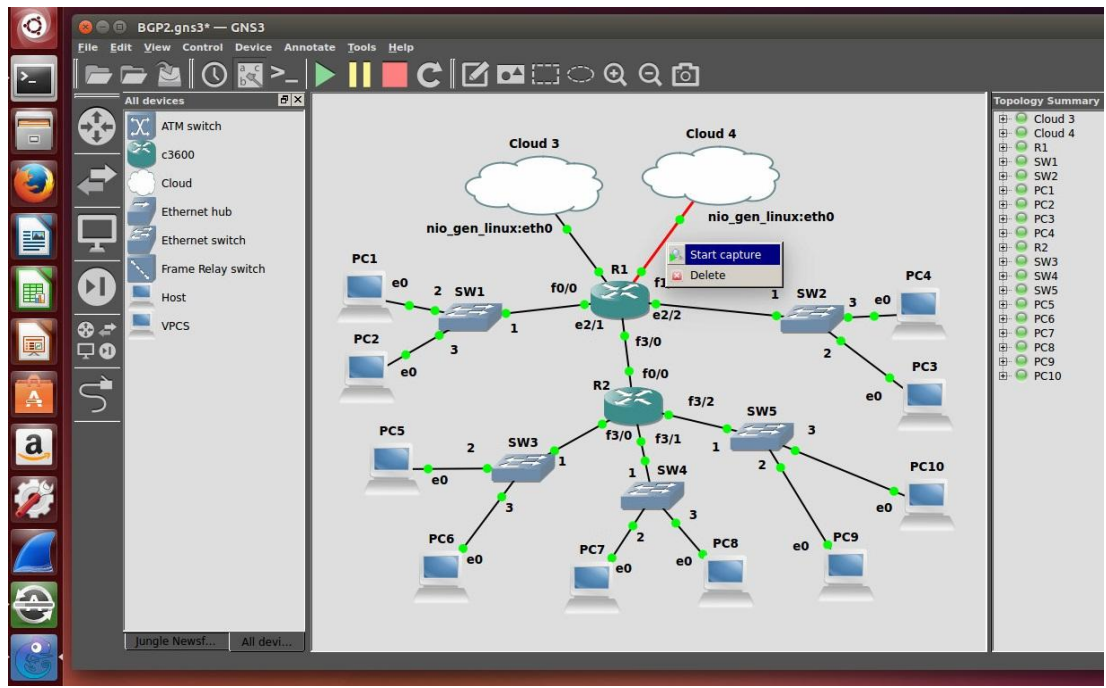

Fig. 6. GNS3 simulation by a student group

## 5. CONCLUSION

High quality hands-on practical exercises are essential for universities to produce work ready graduates in networking. Due to the high cost of networking equipment, it has always been a challenge. This paper presented our effort in designing and building a networking laboratory using both physical equipment and network simulators to offer practical classes for on-campus and online students simultaneously. The network design was centred around the key idea that the networking equipment for each group must be directly connected to a common backbone; and all the laboratory exercises are designed in a way that each group is required to connect to some pre-configured demo networks. The open source network simulator GNS3 provided a low cost solution complementing the existing hardware in the laboratory. We reported our experience in implementing the GNS3 systems, from infrastructure to laboratory exercise design; and also the learning and teaching experiences of using GNS3 in classes. Based on our experience of using GNS3 in an advanced networking subject for one semester, we conclude that the implemented systems meet our expectations and are able to deliver lab exercises effectively for both on-campus and online students. Our next step is to redesign all the existing laboratory exercises for all networking subjects so that they can be delivered on the physical equipment and GNS3 for both on-campus and online students.

**Author**

KC Chan received his BASc in Engineering Science and MASc in Mechanical Engineering from the University of Toronto, Canada, and his PhD in Manufacturing Engineering from the University of New South Wales, Australia. He was previously a tenured Senior Lecturer at the University of New South Wales, and a Visiting Assistant Professor at the Hong Kong University of Science and Technology. Dr. Chan also had extensive industry experience as founders and CTOs in growing multiple technology companies from startups to public companies at the Australian Technology Park. After ten years in the ICT industry and a two years career break, Dr. Chan returned to academia in 2010. He is currently the acting Deputy Head of the Department of Computer Science and Information Technology at La Trobe University, and a Visiting Fellow in Mechatronics at the University of New South Wales. He is serving as the Secretary of the IEEE Computer Society, NSW Chapter. In his academic career, Dr. Chan has published over 80 refereed papers and received over 1.2 million dollars in research funding.